\documentstyle[prl,aps,epsf,floats]{revtex} \def\narrowtext{} \tighten 
\twocolumn
\input epsf.sty

\begin{document}
\draft

\title{High-Resolution Photoemission Study of MgB$_2$}

\author{
        T. Takahashi,$^1$
        T. Sato,$^1$
        S. Souma,$^1$
        T. Muranaka,$^2$
		and J. Akimitsu$^2$
       }
\address{
         (1) Department of Physics, Tohoku University, 980-8578 Sendai, Japan\\
         (2) Department of Physics, Aoyama-Gakuin University, Setagaya-ku, Tokyo 157-8572, Japan\\
         }
\address{%
\begin{minipage}[t]{6.0in}
\begin{abstract}
		 We have performed high-resolution photoemission spectroscopy on MgB$_2$ and observed opening of a superconducting gap
		 with a narrow coherent peak.  We found that the superconducting gap is {\it s}-like with the gap value
		 ($\Delta$) of 4.5$\pm$0.3 meV at 15 K.  The temperature dependence (15 - 40 K) of gap value follows well the BCS form, suggesting that
		 2$\Delta$/{\it k$_B$}{\it T$_c$} at {\it T}=0 is about 3.  No pseudogap behavior is observed in the normal state.  The present results strongly suggest that
		 MgB$_2$ is categorized into a phonon-mediated BCS superconductor in the weak-coupling regime.
\typeout{polish abstract}
\end{abstract}
\pacs{PACS numbers:  74.70.Ad, 74.25.Jb, 79.60.Bm}
\end{minipage}}
\maketitle
\narrowtext
    	The discovery of non-cuprate ``high-temperature" superconductor MgB$_2$ \cite{Nagamatsu} has evoked much attention in how
		this simple binary superconductor is similar to or different from the ``conventional" high-temperature superconductors (HTSCs).
		The superconducting transition temperature ({\it T$_c$}) exceeds those of known metal superconductors and alkali-doped C$_{60}$, 
		being comparable to that of one of HTSCs (La$_{2-x}$Sr$_x$CuO$_4$), and is located just on or a little above the theoretical upper
		limit predicted for the phonon-mediated superconductivity\cite{McMillan}.   It is thus quite important to elucidate the mechanism 
		and origin of this newly discovered ``high-{\it T$_c$}" superconductor in relation to the cuprate HTSCs.  The superconducting gap 
		and its symmetry has been investigated by tunneling \cite{Rubio,Karapetrov,Sharoni} and NMR spectroscopy \cite{Kotegawa}, but the results are not necessarily
		consistent with each other.  Photoemission spectroscopy (PES) has revealed many key physical features of cuprate HTSCs
		such as the pseudo gap \cite{Loeser,Ding} and quasiparticles \cite{Valla,Kaminski}.  A recent remarkable progress in the energy resolution enables direct
		observation of a superconducting gap in metal superconductors \cite{Chainani,Reinert}.  
		
     In this Letter, we report high-resolution PES results on MgB$_2$ to study the superconducting gap and its symmetry. 
	 By using a merit of high energy resolution ($\Delta$E $<$ 5 meV), we have succeeded in directly observing opening/closure 
	 of the superconducting gap as a function of temperature, together with a narrow superconducting coherent peak 
	 located slightly below the Fermi level ({\it E}$_{F}$).  We have performed numerical fittings to the PES spectra to investigate
	 the size and symmetry of superconducting gap.

			The procedure to prepare polycrystalline MgB$_2$ samples has been described elsewhere \cite{Nagamatsu}.  
			The x-ray diffraction measurement shows that the samples are single-phased and the electrical
			resistivity and DC magnetization measurements confirm a sharp superconducting transition onset at
			39.5 K.  High-resolution PES measurements were performed using a SCIENTA SES-200 spectrometer with a
			high-flux discharge lamp and a toroidal grating monochromator.  We set the total energy resolution at 4-7 meV
			to obtain a reasonable count rate near {\it E}$_{F}$.  The base pressure of spectrometer was 1.5x10$^{-11}$ Torr.  Samples were
			scraped {\it in-situ} with a diamond file to obtain a clean surface for PES measurements.  A measured PES spectrum 
			represents the density of states (DOS) since the sample is a polycrystal.   The temperature of sample was monitored
			with a calibrated silicon diode sensor to an accuracy of $\pm$0.5 K.  We have confirmed the results shown here being 
			reproducible for repeated scraping and on a cycle of increasing/decreasing temperature.  The Fermi level of sample 
			was referenced to that of a gold film evaporated onto the sample substrate and its accuracy is estimated to be better 
			than 0.2 meV.

			Figure 1 shows PES spectra near {\it E}$_{F}$ of MgB$_2$ measured with He I$\alpha$ resonance line (21.218 eV) at temperatures (15 and 50 K)
			below/above the {\it T$_c$} (39.5 K).   The spectral intensity is normalized to the area under the curve.  According to the band structure
			calculations, the electronic states near {\it E}$_{F}$ originate in the B 2{\it p} states \cite{Kortus,Belashchenko}.  As seen in Fig. 1, the PES spectrum shows a
			remarkable temperature dependence which is not accounted for by a simple temperature effect due to the Fermi-Dirac function.
			The midpoint of leading edge in the PES spectrum at 15 K is not at {\it E}$_{F}$, but is shifted by a few meV toward the high binding energy
			relative to {\it E}$_{F}$ (see the inset to Fig. 1).  In contrast, the PES spectrum at 50 K appears to have a leading-edge midpoint at {\it E}$_{F}$, 
			though the position is not well defined due to the broad feature of PES spectrum.  It is remarked here that the 15-K spectrum 
			has a ``peak" structure around 10 meV, which totally disappears at 50 K.  A similar peak structure has been observed in cuprate 
			HTSCs such as Bi$_2$Sr$_2$CaCu$_2$O$_8$, where the peak is located at much higher binding energy (about 40 meV) and is ascribed to the
			superconducting coherent peak \cite{Mesot}.  Thus the present temperature-dependent PES results unambiguously show that a 
			superconducting gap of a few meV opens at {\it E}$_{F}$ in MgB$_2$ at low temperatures.  The observed temperature-induced change 
			of PES spectrum is understood in terms of ``pile-up" of the electronic states transferred from the near-{\it E}$_{F}$ region at the 
			superconducting state.
		
\begin{figure}[!t]
\epsfxsize=3.4in
\epsfbox{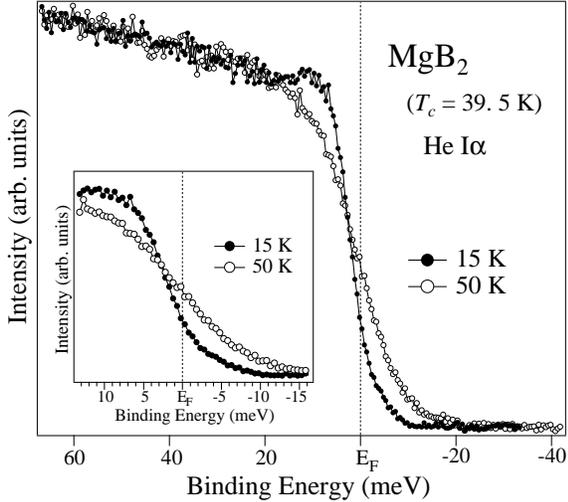}
\vspace{0.1cm}
\caption{
High-resolution PES spectra near {\it E}$_{F}$ of MgB$_2$ polycrystal measured with He I$\alpha$ resonance line (21.218 eV) at temperatures 
(15 and 50 K)  below/above the {\it T$_c$} (39.5 K).   The spectral intensity is normalized to the area under the curve.  
The inset shows the expansion near {\it E}$_{F}$.}
\label{fig1}
\end{figure}

		          The next problem to be solved is the size and symmetry of superconducting gap.  Figure 2 shows comparison of the
				  PES spectra of MgB$_2$ with those of gold measured under the same experimental condition.  The leading edge of MgB$_2$
				  spectrum at 15 K is shifted toward the high binding energy with respect to the gold reference, indicative of the gap 
				  opening at low temperature.  On the other hand,  the spectrum at 50 K looks to coincide with the gold reference within 
				  the present experimental accuracy, indicating that the gap is closed at 50 K with no pseudogap in contrast to the
				  underdoped cuprate HTSCs \cite{Loeser,Ding}.  We find that the leading edge of MgB$_2$ spectrum at 15 K is almost parallel to that of
				  gold reference, which suggests that the symmetry of gap is {\it s}-like and the gap size is within a few meV.  In order to
				  confirm this, we have performed numerical fittings to the PES spectrum and show the results in Fig. 3.  In the simulation, 
				  we used the Dynes function \cite{Dynes} multiplied by the Fermi-Dirac function at {\it T}=15 K and convoluted it with a Gaussian 
				  with a full-width-at-half-maximum (FWHM) of the instrumental resolution.  We set the broadening parameter $\Gamma$ to 1.1 meV.  As shown in Fig. 3, the fittings with
				  {\it d}-wave gap of $\Delta$=5-10 meV seem to hardly reproduce the experimental PES spectrum at 15 K; both the coherent peak 
				  and the leading edge are not well reproduced simultaneously.  On the other hand, the simulation with {\it s}-wave gap of 
				  $\Delta$=4.5 meV looks to fairly well reproduce the experimental curve.  We found that the change of gap value by 0.5 meV 
				  in the fittings leads to an apparent deviation from the experimental spectrum.  It is thus most probable that MgB$_2$ has a {\it s}-like 
				  superconducting gap and the gap value at 15 K is 4.5$\pm$0.3 meV.   The inset to Fig. 3 shows the gap value as a function 
				  of temperature obtained from the numerical fittings to the PES spectra measured at several temperature of 15 - 40 K.
				  The temperature dependence follows well the BCS form, suggesting the gap value at {\it T}=0 being about 5 meV.  
				  This gap value gives the ratio 2$\Delta$/{\it k$_B$}{\it T$_c$} of about 3, which is slightly smaller than the theoretical value (3.53) for a 
				  weak-coupling BCS superconductor \cite{Bardeen}.   All these strongly suggest that MgB$_2$ is categorized into a phonon-mediated 
				  BCS superconductor in the weak-coupling regime.  It is inferred that high-frequency phonon(s) due to the light 
				  mass of boron atoms may cause the ``high-temperature superconductivity" in MgB$_2$.
				  
			\begin{figure}[!t]
\epsfxsize=3.4in
\epsfbox{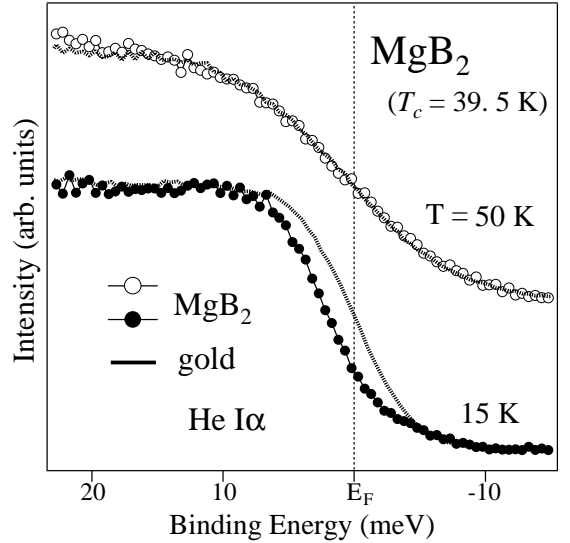}
\vspace{0.1cm}
\caption{
Comparison of high-resolution PES spectra of MgB$_2$ with those of gold measured under the same experimental condition.
}
\label{fig2}
\end{figure}

\begin{figure}[!t]
\epsfxsize=3.4in
\epsfbox{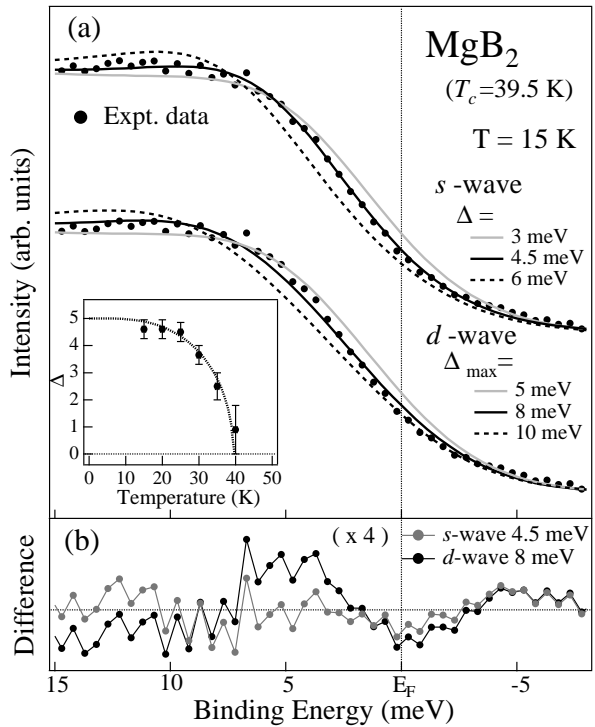}
\vspace{0.1cm}
\caption{
(a) Results of numerical fittings to the high-resolution PES spectrum of MgB$_2$ at 15 K.  The experimental spectrum is shown by filled dots
 and the fitting curves by lines.  The inset shows the temperature dependence of the gap value obtained from the fittings to the PES 
 spectra at several temperatures of 15 - 40 K.  (b) Difference curves between the experimental spectrum and the representative 
numerical fitting curves based on a {\it s}- or {\it d}-wave gap.}
\label{fig3}
\end{figure}
			
	      In conclusion, we have performed high-resolution photoemission spectroscopy on a newly discovered novel superconductor MgB$_2$. 
		  We have clearly observed opening/closure of a superconducting gap at {\it E}$_{F}$ on decreasing/increasing temperature across {\it T$_c$}. 
		  The gap is found to be {\it s}-like and the gap value ($\Delta$) is estimated to be 4.5$\pm$0.3 meV at 15 K from the numerical fittings to the
		  PES spectra. The temperature dependence of gap value is well described with the BCS form.  The present PES results indicate
		  that MgB$_2$ is categorized into a phonon-mediated BCS superconductor in the weak-coupling regime.\\

			We thank T. Kamiyama, S. Nishina, and H. Matsui for their help in PES measurements.  
			This work was supported by grants from the CREST (Core Research for Evolutional 
			Science and Technology Corporation) of JST, the Japan Society for Promotion of Science (JSPS),
			and the Ministry of Education, Science and Culture of Japan. TS thanks the JSPS for financial support.


\begin{references}
\bibitem{Nagamatsu}
J. Nagamatsu {\it et al.}, Nature (London) {\bf 410}, 63, (2001).
\bibitem{McMillan}
W. L. McMillan, Phys. Rev. {\bf 167}, 331 (1968).
\bibitem{Rubio}
G. Rubio-Bollinger, H. Suderow, S. Vieira, cond-mat /0102242.
\bibitem{Karapetrov}
G. Karapetrov {\it et al.}, cond-mat /0102312.
\bibitem{Sharoni}
A. Sharoni, I. Felner, and O. Millo, cond-mat /0102325.
\bibitem{Kotegawa}
H. Kotegawa {\it et al.}, cond-mat/0102334.
\bibitem{Loeser}
A. G. Loeser {\it et al.}, Science {\bf 273}, 325 (1996).
\bibitem{Ding}
H. Ding {\it et al.}, Nature (London) {\bf 382}, 51 (1996).
\bibitem{Valla}
T. Valla {\it et al.}, Science {\bf 285}, 2110 (1999).
\bibitem{Kaminski}
A. Kaminski {\it et al.}, Phys. Rev. Lett. {\bf 84}, 1788 (2000).
\bibitem{Chainani}
A. Chainani {\it et al.}, Phys. Rev. Lett. {\bf 85}, 1966 (2000).
\bibitem{Reinert}
F. Reinert {\it et al.}, Phys. Rev. Lett. {\bf 85}, 3930 (2000).
\bibitem{Kortus}
J. Kortus {\it et al.}, cond-mat/0101446.
\bibitem{Belashchenko}
K. D. Belashchenko, M. van Schilfgaarde and V. P. Antropov, cond-mat /0102290,
\bibitem{Mesot}
J. Mesot {\it et al.}, Phys. Rev. Lett. {\bf 83}, 840 (1999).
\bibitem{Dynes}
R. C. Dynes, V. Narayanamurti, and J. P. Garno, Phys. Rev. Lett. {\bf 41}, 1509 (1965).
\bibitem{Bardeen}
J. Bardeen, L. N. Cooper, and J. R. Schrieffer, Phys. Rev. {\bf 108}, 1175 (1957).
\end{references}
\end{document}